\documentclass[twocolumn,superscriptaddress,floats,prd,nofootinbib]{revtex4}

\usepackage{amssymb}
\usepackage{stmaryrd}
\usepackage{amsmath}
\usepackage{amsfonts}
\usepackage{mathrsfs}
\usepackage{CJK}
\usepackage{amsmath,amssymb,amsfonts}
\usepackage{graphicx}
\usepackage{subfigure}
\usepackage{appendix}

\usepackage[T1]{fontenc}
\usepackage{xcolor}
\usepackage{ulem}

\def\be{\begin{equation}}
\def\ee{\end{equation}}
\def\ba{\begin{eqnarray}}
\def\ea{\end{eqnarray}}
\def\nn{\nonumber}

\newcommand{\mubar}{{\bar \mu}} 
\newcommand{\abs}[1]{{\left|{#1}\right|}} 
\newcommand{\Abs}[1]{{\Big|{#1}\Big|}} 
\newcommand{\ket}[1]{\vert{#1}\rangle} 
\newcommand{\R}{\mathcal {R}} 
\newcommand{\ct}{\tilde{c}}
\newcommand{\ints}{{\int_\Sigma}} 

\newcommand{\grav}{\mathrm{gr}} 
\newcommand{\sca}{\mathrm{sc}} 
\newcommand{\kin}{\mathrm{kin}} 
\newcommand{\hil}{\mathcal{H}} 

\begin{document}

\begin{titlepage}
\title{Loop quantum gravity and cosmological constant }
\author{Xiangdong Zhang}
\thanks{scxdzhang@scut.edu.cn}
\affiliation{Department of Physics, South China University of Technology, Guangzhou 510641, China}
\author{Gaoping Long}
\thanks{201731140005@mail.bnu.edu.cn}
\affiliation{Department of Physics, South China University of Technology, Guangzhou 510641, China}
\author{Yongge Ma}
\thanks{Corresponding author. mayg@bnu.edu.cn}
\affiliation{Department of Physics, Beijing Normal University, Beijing 100875, China}

\begin{abstract}
An one-parameter regularization freedom of the Hamiltonian constraint for loop quantum gravity is analyzed. The corresponding spatially flat, homogenous and isotropic model includes the two well-known models of loop quantum cosmology as special cases. The quantum bounce nature is tenable in the generalized cases. For positive value of the regularization parameter, the effective Hamiltonian leads to an asymptotic de-Sitter branch of the Universe connecting to the standard Friedmann branch by the quantum bounce. Remarkably, by suitably choosing the value of the regularization parameter, the observational cosmological constant can emerge at large volume limit from the effect of quantum gravity, and the effective Newtonian constant satisfies the experimental restrictions in the meantime.
\end{abstract}


\maketitle

\end{titlepage}

\section{Introduction}

The origin of current cosmic acceleration is one of the biggest challenges to modern physics, which is usually called as the dark energy issue. Many possible mechanisms have been proposed to account for this issue, such as the phenomenological models\cite{Friemann08}, modified gravity \cite{Banerjee,Sen,Qiang,Peebles03}, higher dimensions \cite{Tye01} and so on. Among them, the cosmological constant is generally believed as the most simplest explanation \cite{Peebles03,Weinberg89}. However, the nature of the cosmological constant is still mysterious. Whether it a purely classical effect or it has a quantum origin is a crucial open issue. It is well known that the awkward cosmological constant problem would appear if one considered quantum matter fields on a classical spacetime background \cite{Peebles03,Weinberg89}. A challenging question would be whether a realistic cosmological constant could emerge from certain quantum gravity theory.

How to unify general relativity(GR) with quantum mechanics remains as the biggest theoretical challenge to fundamental physics. Among various approaches to quantum gravity, loop quantum gravity (LQG) is notable with its background independence \cite{Ro04,Th07,As04,Ma07}.
The non-perturbative quantization procedure of LQG has
been applied not only to GR, but also to the metric $f(\R)$ theories\cite{Zh11,Zh11b},
scalar-tensor theories\cite{ZM12a,ZM11c}, and higher dimensional gravity \cite{Thiemann13}. The idea and technique of LQG have been successfully carried out in the
symmetry-reduced models of loop quantum cosmology (LQC). We refer to \cite{LQC5,Boj,APS3,AS11} for reviews on LQC.

A remarkable result of LQC is that the classical big bang singularity of the Universe can be avoided by a quantum bounce \cite{LQC5,Boj,APS3,AS11,YDM,DMY}. Moreover, LQC opens
a promising avenue to relate quantum gravity effects to cosmological observations of the very early Universe
\cite{Agullo12,AG}. As in any quantization procedure of a classical theory, different regularization schemes exist also in LQC as well as in LQG \cite{Th07,As04,YM15,Lewandowski15,Singh18}. In particular, for the LQC model of flat Friedmann-Lemaitre-Robertson-Walker (FLRW) universe, alternative Hamiltonian constraint operators were proposed \cite{APS,YDM}. Note that, to inherit more features from LQG, the so-called Euclidean term and Lorentzian term of the Hamiltonian constraint were treated independently in the LQC model in \cite{YDM}. It was recently shown in \cite{Pawlowski18,Pawlowski19} that one of the quantum Hamiltonians proposed in \cite{YDM} can lead to a
new evolution scenario where the prebounce
geometry could be described at the effective level by a de
Sitter spacetime. This raises the possibility to obtain a positive cosmological constant from a model of LQC.  However, the cosmological
constant obtained in \cite{Pawlowski18} is very large ($\Lambda\approx 1.03\ell^{-2}_{Pl}\sim10^{70}m^{-2}$) and thus fails in fitting the current observations which requires a very small cosmological constant ($\Lambda_{ob}\sim 1.09168\times 10^{-52}m^{-2}$). Fortunately, this is not the end story. In this Letter, we will reveal the new possibility that even if one started with the classical GR without cosmological constant, there still exists certain regularization of the Hamiltonian in LQC such that a small enough positive cosmological constant can emerge at the effective level. This is a quantum dynamical effect significantly different form the usual scenario where one could
add a non-dynamical cosmological constant to both classical and quantum Einstein's equations and fix its value by observations. Moreover, the regularization choice inherits that of full LQG. It is reasonable to infer that the effective Hamiltonian could also be obtained by suitable semiclassical analysis of certain Hamiltonian of LQG.

\section{Classical setting}
LQG is based on the connection-dynamical formulation of GR defined on a spacetime
manifold $M=R\times \Sigma$, where $\Sigma$ denotes
a three-dimensional spatial manifold. The classical phase space consists of the
Ashtekar-Barbero variables $(A_a^i(x),E_i^a(x))$, where $A_a^i(x)$ is a $SU(2)$ connection and $E^i_a(x)$ is a
densitized triad \cite{As04,Th07,Ma07}. The non-vanishing Poisson
bracket is given by
\ba
\{A_a^i(x),E_j^b(y)\}=8\pi G\gamma\delta_a^b\delta_j^i\delta^3(x,y)
\ea where $G$ is the gravitational constant and $\gamma$ is the Barbero-Immirzi parameter. The
classical dynamics of GR is thus obtained by imposing the Gaussian, diffeomorphism and
Hamiltonian constraints on the phase space, where the latter represents the reparametrization freedom of time variable. In LQG, the notable Hamiltonian constraint operator proposed in \cite{Thiemann98} and an alternative one proposed in \cite{YM15} are based on the regularization schemes of the following expression of the Hamiltonian
constraint
{\footnotesize\ba
H_{g}=\frac{1}{16\pi G}\ints d^3xN\left[F^j_{ab}-(\gamma^2+1)\varepsilon_{jmn}K^m_aK^n_b\right]\frac{\varepsilon_{jkl}
E^a_kE^b_l}{\sqrt{q}},\nn\\\label{Hamiltoniang}
\ea}where $N$ is the lapse function, $q$ denotes the determinant  of the spatial metric, $F^i_{ab}$ is the curvature of connection $A^i_a$, and $K^i_a$ represents
the extrinsic curvature of $\Sigma$.  The so-called Euclidean term $H^E$ and Lorentzian term $H^L$ in Eq. \eqref{Hamiltoniang} are denoted respectively as
\ba
H^E&=&\frac{1}{16\pi G}\ints d^3 x N F^j_{ab}\frac{\varepsilon_{jkl}
E^a_kE^b_l}{\sqrt{q}},\label{HamiltonianE}
\ea and
\ba
H^L&=&\frac{1}{16\pi G}\ints d^3xN\left(\varepsilon_{jmn}K^m_aK^n_b\right)\frac{\varepsilon_{jkl}
E^a_kE^b_l}{\sqrt{q}}.\label{HamiltonianL}
\ea
There is another alternative Hamiltonian constraint operator proposed in \cite{Lewandowski15} for LQG, which is based on the regularization scheme of the following expression of the Hamiltonian constraint,
{\small\ba
H_{g}&=&-\frac{1}{16\pi G \gamma^2}\ints d^3xN\left[F^j_{ab}\frac{\varepsilon_{jkl}
E^a_kE^b_l}{\sqrt{q}}+(\gamma^2+1)\sqrt{q}R\right]\label{HamiltoniancE}\nn\\
\ea}where $R$ is the 3-dimensional spatial curvature of $\Sigma$.
It is easy to see that expressions \eqref{Hamiltoniang} and \eqref{HamiltoniancE} are equivalent to each other by using the classical identity (up to Gaussian constraint)\cite{Th07} \ba
H^E=\gamma^2H^L-\ints\sqrt{q}R.
\ea Here, we point out that there is an one-parameter freedom to express the classical Hamiltonian constraint in the connection formalism. The general expression can be written as
\ba
H_{g}&=&\lambda H^E-(1+\lambda\gamma^2)H^L+(-1+\lambda)\ints\sqrt{q}R,\label{generalH}
\ea
where $\lambda$ is an arbitrary real number to represent the freedom of choices. Clearly, the expression \eqref{Hamiltoniang} corresponds to the choice of $\lambda=1$, while the expression \eqref{HamiltoniancE} corresponds to the case of $\lambda=-\frac{1}{\gamma^2}$. It should be noted that all the classical theories corresponding to the different choices of $\lambda$ are equivalent to each other. However, the quantization of classically equivalent expressions could lead to nonequivalent operators. Therefore, different choices of $\lambda$ might correspond to different quantum theories. This is the case for the LQC model which we are going to consider. Our idea is to use experiments or observations to single out the preferred expression of the Hamiltonian (or the free parameter $\lambda$ ).

Now we consider the spatially flat FLRW model. One has to introduce an ``elemental cell" $\mathcal {V}$ on the spatial manifold $\mathbb{R}^3$ and restrict all
integrals to this cell. Then one chooses a fiducial Euclidean metric ${}^oq_{ab}$ on $\mathbb{R}^3$, as well as the orthonormal triad and co-triad
$({}^oe^a_i ; {}^o\omega^i_a)$ adapted to $\mathcal {V}$ such that
${}^oq_{ab}={}^o\omega^i_a{}^o\omega^i_b$.
Via fixing the degrees of freedom of local gauge and
diffeomorphism transformations, one can obtain the reduced connection and densitized
triad as \cite{LQC5}
\ba A_a^i=\ct
V_0^{-\frac13}\ {}^o\omega^i_a,\quad\quad\quad
E^b_j=pV_0^{-\frac23}\sqrt{\det({}^0q)}\ {}^oe^b_j, \nn\ea
where $V_o$ is the volume of $\mathcal {V}$ measured by
${}^oq_{ab}$, $\ct,p$
are only functions of the cosmological time $t$. To identify a dynamical matter field as an internal clock, we employ a massless scalar field $\phi$ with Hamiltonian \ba
H_\phi=\frac{p^2_\phi}{2\abs{p}^{\frac32}},
\ea where $p_\phi$ is the momentum of $\phi$. Hence the phase space of
the cosmological model consists of conjugate pairs $(\ct,p)$ and
$(\phi,p_\phi)$, with the following nontrivial Poisson brackets,
\ba
\{\ct,p\}&=&\frac{8\pi G}{3}\gamma,\quad\{\phi,p_\phi\}=1. \label{poissonb}
\ea
Note that the gravitational variables are related to the scale factor $a$ by $\abs{p}=a^2V_0^{\frac 23}$ and
$\ct=\gamma\dot{a} V_0^{\frac 13}$. Since all the Hamiltonians corresponding to \eqref{generalH} with different  choices of $\lambda$ are equivalent to each other, they all lead to the standard Friedman equation without cosmological constant as \ba
H^2=\left(\frac{\dot{a}}{a}\right)^2=\frac{8\pi G}{3}\rho.
\ea

\section{Quantum dynamics}

In the physically reasonable $\mubar$ scheme of LQC \cite{APS3}, it is convenient to introduce new conjugate variables for gravity by the canonical transformation
\ba v:=2\sqrt{3}sgn(p)\mubar^{-3},\quad b:=\mubar \ct, \nn\ea
where
$\mubar=\sqrt{\frac{\Delta}{|p|}}$ with
$\Delta=4\sqrt{3}\pi\gamma G\hbar$ being the minimum
nonzero eigenvalue of the area operator \cite{Ash-view}. The new variables also
form a pair of conjugate variables as
\ba \{b,v\}=\frac{2}{\hbar}\ .\nn
\ea
In LQC, the
kinematical Hilbert space for the geometry part is defined as
$\mathcal{H}_{\kin}^{\grav}:=L^2(R_{Bohr},d\mu_{H})$, where
$R_{Bohr}$ and $d\mu_{H}$ are respectively the Bohr
compactification of the real line and Haar measure on it
\cite{LQC5}. The kinematical Hilbert space for the scalar field part is defined as
in usual Schrodinger representation by
$\mathcal{H}_{\kin}^{\sca}:=L^2(R,d\mu)$. Hence the whole Hilbert
space of our model is a direct product, $\mathcal{H}_{\kin}=\mathcal{H}^{\grav}_{\kin}\otimes\mathcal{H}^{\sca}_{\kin}$. In $\mathcal{H}^{\grav}_{\kin}$, there are two elementary operators, $\widehat{e^{ib/2}}$ and $\hat{v}$.
It turns out that the eigenstates $\ket{v}$ of
 $\hat{v}$ contribute an orthonormal basis in $\mathcal{H}_{\kin}^{\grav}$. In the $v$-representation, the actions of  these two operators on the basis read \ba
 \widehat{e^{\frac{ib}{2}}}\ket{v}&=&\ket{v+1},\quad \hat{v}\ket{v}=v\ket{v}.
 \ea
Let $|\phi)$ be the generalized eigenstates of $\hat{\phi}$ in $\hil^\sca_\kin$. We denote
$|\phi,v):=\ket{v}\otimes|\phi)$ as the generalized basis for $\hil_\kin$.

Notice that the spatial curvature $R$ vanishes in the spatially flat FLRW model. Hence the general expression \eqref{generalH} reduces to
{\footnotesize\ba
H_{g}=\frac{1}{16\pi G}\int d^3x\left[\lambda F^j_{ab}-(\lambda\gamma^2+1)\varepsilon_{jmn}K^m_aK^n_b\right]\frac{\varepsilon_{jkl}
E^a_kE^b_l}{\sqrt{q}}\label{Hamiltoniana}\nn\\
\ea}By the regularization procedure mimicking that in full LQG, both the Euclidean term $H^E$ \cite{APS3} and the Lorentzian term $H^L$ \cite{YDM} have been quantized as well-defined operators in $\mathcal{H}_{\kin}^{\grav}$. Therefore the operators $\hat{H}_G$ corresponding to \eqref{Hamiltoniana} is ready.
Its action on a wave function $\Psi(v,\phi)$ in $\hil_\kin$ is the following difference equation,
\ba
\hat{H}_G\Psi(v,\phi)&=&f_+(v)\Psi(v+4,\phi)+(f_0(v)+g_0(v))\Psi(v,\phi)\nn\\
&+&f_-(v)\Psi(v-4,\phi)+g_+(v)\Psi(v+8,\phi)\nn\\
&+&g_-(v)\Psi(v-8,\phi),
\ea
where
{\small\ba
f_+(v)&=-\frac{27\lambda}{16}\sqrt{\frac{8\pi}{6}}\frac{\sqrt{G\hbar}}{8\pi G\gamma^{\frac32}}\Abs{\abs{v+2}-\abs{v}}\abs{v+1},\nn\\
f_-(v)&=f_+(v-2), \quad  f_0(v)=-f_+(v)-f_-(v).\nn\\
g_+(v)&=-\frac{(1+\lambda\gamma^2)\sqrt{6}}{2^{8}\times 3^3}\,\frac{\gamma^{3/2}}{(8\pi G)^{3/2}\hbar^{1/2}}\,\frac{1}{L}\tilde{g}_+(v),\nn\\
\tilde{g}_+(v)&=\Big[M_v(1,5)f_+(v+1)-
     M_v(-1,3)f_+(v-1)\Big]\nonumber\\
       &\quad\times(v+4)M_v(3,5)\nonumber\\
       &\quad\times\Big[M_v(5,9)f_+(v+5)-M_v(3,7)f_+(v+3)\Big],\nonumber\\
g_-(v)&=-\frac{(1+\lambda\gamma^2)\sqrt{6}}{2^{8}\times 3^3}\,\frac{\gamma^{3/2}}{(8\pi G)^{3/2}\hbar^{1/2}}\,\frac{1}{L}\tilde{g}_-(v),\nn\\
\tilde{g}_-(v)&=\Big[M_v(1,-3)f_-(v+1)-M_v(-1,-5)f_-(v-1)\Big]\nonumber\\
       &\quad\times(v-4)M_v(-5,-3)\nonumber\\
       &\quad\times\big[M_v(-3,-7)f_-(v-3)-M_v(-5,-9)f_-(v-5)\big],\nonumber\\
g_o(v)&=-\frac{(1+\lambda\gamma^2)\sqrt{6}}{2^{8}\times 3^3}\,\frac{\gamma^{3/2}}{(8\pi G)^{3/2}\hbar^{1/2}}\,\frac{1}{L}\tilde{g}_o(v),\nn\\
\tilde{g}_o(v)&=\Big[M_v(1,5)f_+(v+1)-M_v(-1,3)f_+(v-1)\Big]\nonumber\\
       &\quad\times(v+4)M_v(3,5)\nonumber\\
       &\quad\times\Big[M_v(5,1)f_-(v+5)-M_v(3,-1)f_-(v+3)\Big]\nonumber\\
       &+\Big[M_v(1,-3)f_-(v+1)-M_v(-1,-5)f_-(v-1)\Big]\nonumber\\
       &\quad\times(v-4)M_v(-5,-3)\nonumber\\
       &\quad\times\Big[M_v(-3,1)f_+(v-3)-M_v(-5,-1)f_+(v-5)\big],\nn
\ea}with $M_v(a,b):=|v+a|-|v+b|$ and $L=\frac{4}{3}\sqrt{\frac{\pi\gamma G\hbar}{3\Delta}}$. Thus, the Hamiltonian constraint equation of our LQC model can be written as
\ba
\left(\hat{H}_{G}+\frac{\sqrt{3}\hat{p}_\phi^2}{(\Delta)^{\frac32}}\widehat{\abs{v}^{-1}}\right)\Psi(\phi,v)=0,\label{hbd}
\ea
where the action of the Hamiltonian of matter field reads
\ba
\frac{\sqrt{3}\hat{p}_\phi^2}{(\Delta)^{\frac32}}\widehat{\abs{v}^{-1}}\Psi(v,\phi)=-\frac{\sqrt{3}}{(\Delta)^{\frac32}}\hbar^2B(v)
\frac{\partial^2\Psi(v,\phi)}{\partial\phi^2},\label{hm} \ea
with $B(v)=\left(\frac{3}{2}\right)^{3}\abs{v}\abs{\abs{v+1}^{1/3}-\abs{v-1}^{1/3}}^3$ \cite{APS3}. Note that we still have the freedom to choose a particular value of the parameter $\lambda$. It is obvious that, if one set $\lambda=-\frac{1}{\gamma^2}$, Eq.\eqref{hbd} would coincide with the quantum dynamics in \cite{APS3}, while by choosing $\lambda=1$, one of the Hamiltonians in \cite{YDM} would be obtained. Our viewpoint is that the value of $\lambda$ should be fixed by observations.
To this aim, let us study the effective
theory indicated by Eq.\eqref{hbd}. It has been showed in \cite{YDM} that the expectation values of the Euclidean term $H^E$ and the Lorentzian term $H^L$ at sub-leading order read respectively as
\ba
\langle \widehat{H}^E\rangle&=\frac{3\abs{v}\beta}{8\pi G\Delta}\left[e^{-4\epsilon^2}\sin^2(b)+\frac12(1-e^{-4\epsilon^2})\right],\nn\\
\langle \widehat{H}^L\rangle&=\frac{3\abs{v}\beta}{32\pi G\gamma^2\Delta}\left[e^{-16\epsilon^2}\sin^2(2b)+\frac12(1-e^{-16\epsilon^2})\right]\label{expectationH}
\ea where $\beta=2\pi G\hbar\gamma\sqrt{\Delta}$ and $\epsilon=\frac{1}{d}$ with $d$ denoting the characteristic "width" of the semiclassical state. Thus the total effective Hamiltonian
constraint of the model at leading order reads
{\small\ba
H_F=-\frac{3\beta}{8\pi G\gamma^2\Delta}|v|\sin^2b\left(1-(1+\lambda\gamma^2)\sin^2(b)\right)+\beta|v|\rho\label{effectiveH},
\ea}
where $\rho=\frac{p^2_\phi}{2V^2}$ with $V=\abs{p}^{\frac32}$ being the physical volume of $\mathcal{V}$. It should be noted that although Eq. \eqref{effectiveH} could be obtained from the Hamiltonian studied in Refs. \cite{YDM,Pawlowski18} after doing the rescalings $\gamma^2 \mapsto\lambda \gamma^2$ and $\Delta \mapsto\Delta/\lambda$, the theory is not invariant under the rescalings. The Hamiltonian \eqref{effectiveH} represents a family of effective theories beyond that in Refs. \cite{YDM,Pawlowski18}.

Now we consider the effective Hamiltonian \eqref{effectiveH} with $\lambda>0$. At the kinematical level, the matter energy-density $\rho$ can be solved by the effective Hamiltonian constraint $H_F=0$ as
\begin{equation}\label{matterdensity}
\rho=\frac{3}{8\pi G\Delta \gamma^2 }\sin^2b(1-(1+\lambda\gamma^2)\sin^2b).
\end{equation}
This in turn implies two solutions $b_+$ and $b_-$ satisfying
\ba
\sin^2(b_{\pm})=\frac{1\pm \sqrt{1-\frac{\rho}{\rho_{c}}}}{2(1+\lambda\gamma^2)}\label{bpm},
\ea  where $\rho_c=\frac{3}{32\pi G(1+\lambda\gamma^2)\gamma^2\Delta}$. Takeing into account the fact that $0<\sin^2b\leq\frac{1}{1+\lambda\gamma^2}$, it is easy to see that $\rho$ is bounded by its maximum value $
\rho_c$. The effective equations of motion of the model with respect to the cosmological time $t$ can be derived by the Hamiltonian constraint \eqref{effectiveH}. In particular, one can obtain
\ba\label{eom2}
\dot{v}&=\frac{3\beta}{4\pi G\hbar \gamma^2{\Delta} }|v|\sin(2b)(1-2(1+\lambda\gamma^2)\sin^2b),\\
\dot{b}&= -\frac{ p_\phi^2}{\hbar\beta v^2}-\frac{3\beta}{4\pi G\hbar \gamma^2{\Delta} }\sin^2b(1-(1+\lambda\gamma^2)\sin^2b).
\ea
Hence we have $\dot{v}=0$ for $b_c$ satisfying $\sin^2(b_c)=\frac{1}{2(1+\lambda\gamma^2)}$. Moreover, the second order derivative of $v$ can be calculated at this point as
{\small\ba
\ddot{v}|_{b=b_c}&=&24\pi G(1+\beta^2)\beta^2|v|\sin^2(2b_0)(1+\lambda\gamma^2)\rho|_{b=b_c}>0\nn
\ea}Hence the matter density $\rho$ takes its maximum at the bounce point. Therefore the point $v|_{b=b_c}$ is the minimum where the quantum bounce of the Universe happens. In terms of the scale factor $a$, the Friedmann and Raychaudhuri equations of this model are derived as
{\small \ba\label{Fri}
H^2
&=&\frac{1}{\gamma^2\Delta}\sin^2(b)(1-\sin^2(b))(1-2(1+\lambda\gamma^2)\sin^2b)^2\nn\\
\ea
\begin{eqnarray}\label{Ray}
\frac{\ddot{a}}{a}&=&(H)^2 +\frac{1}{\gamma\sqrt{\Delta}}\dot{b}( 1-2\sin^2(b)\nn\\
&&-2(1+\lambda\gamma^2)\sin^2b(3-4\sin^2b)),
\end{eqnarray}}where $H$ denotes the Hubble parameter and $b$ has two solutions as shown in \eqref{bpm}. We are going to show that the two solutions correspond to the two periods of the evolution of the Universe, divided by the bounce point.

The Hamiltonian \eqref{effectiveH} implies that $p_\phi$ is constant of motion and hence $\phi$ is monotonic with respect to $t$. By identifying $\phi$ as a dynamical time, we obtain the following analytic solution of the effective equations of motion in the case of $\lambda>0$,
{\small \begin{equation}\label{xphi}
x(\phi)=\frac{1}{1+\lambda\gamma^2\cosh^2(\sqrt{\frac{3\beta^2}{\hbar^2\pi G\Delta \gamma^2 }}(\phi-\phi_o))},
\end{equation}
\begin{equation}
\rho(\phi)=\frac{3\lambda}{8\pi G \Delta}(\frac{\sinh(\sqrt{\frac{3\beta^2}{\hbar^2\pi G\Delta \gamma^2 }}(\phi-\phi_o))}{1+\lambda\gamma^2\cosh^2(\sqrt{\frac{3\beta^2}{\hbar^2\pi G\Delta \gamma^2 }}(\phi-\phi_o))})^2
\end{equation}
and
\begin{equation}\label{vphi}
v(\phi)=\sqrt{\frac{ 4\pi G |p_\phi|^2\Delta}{3\lambda\beta^2}}\frac{1+\lambda\gamma^2\cosh^2(\sqrt{\frac{3\beta^2}{\hbar^2\pi G\Delta \gamma^2 }}(\phi-\phi_o))}{|\sinh(\sqrt{\frac{3\beta^2}{\hbar^2\pi G\Delta \gamma^2 }}(\phi-\phi_o))|},
\end{equation}}where we defined $x:=\sin^2b$, and $\phi_o$ is an integral constant. Eq.\eqref{xphi} indicates that the range of $x$ in the physical evolution covers the interval $(0,\frac{1}{1+\lambda\gamma^2})$. This confirms that the bounce point of $x= \frac{1}{2(1+\lambda\gamma^2)}$ does appear in the physical evolution. Moreover, by using the Hamiltonian \eqref{effectiveH} and Eq.\eqref{vphi}, the relation between $\phi$ and $t$ can be solved as
{\scriptsize\begin{widetext}
\begin{equation}
t(\phi)-t_0=\frac{2\pi G\hbar\Delta\gamma^3\sqrt{\lambda}\text{sgn}[p_\phi(\phi-\phi_0)]}{3\beta}\left(\cosh(\sqrt{\frac{3\beta^2}{\hbar^2\pi G\Delta \gamma^2 }}(\phi-\phi_0))-\frac{1+\lambda\gamma^2}{\lambda\gamma^2}\ln|\coth(\sqrt{\frac{3\beta^2}{4\pi G\hbar^2\Delta \gamma^2 }}(\phi-\phi_0))|\right),\nn
\end{equation}
\end{widetext}}where $\text{sgn}[p_\phi(\phi-\phi_0)]$ denotes the sign of $p_\phi(\phi-\phi_0)$. Therefore, either of the two domains $\phi>\phi_0$ and $\phi<\phi_0$ can cover the range of $t$. We thus consider the domain $\phi>\phi_0$ without loss of generality. Then the infinite past and infinite future of $t$ correspond to $\phi\to\phi_0^+$ and $\phi\to+\infty$ respectively. Hence Eq.\eqref{xphi} ensures that the two solutions $b_-$ and $b_+$ in \eqref{bpm} correspond to the two periods given by $0<\sin^2b_-\leq \frac{1}{2(1+\lambda\gamma^2)}$ and $\frac{1}{2(1+\lambda\gamma^2)}\leq\sin^2b_+< \frac{1}{1+\lambda\gamma^2}$ of the evolution of the Universe, divided by the bounce point at $b_c$. By Eqs. \eqref{vphi} and \eqref{xphi}, it is obvious that $v\rightarrow\infty$ is achieved at $(\phi-\phi_o)\rightarrow+\infty$ or $(\phi-\phi_o)\rightarrow0^+$, and correspondingly the behaviour of $b$ is given by
\ba
b &\rightarrow& 0 \quad \text{if}\  (\phi-\phi_o)\rightarrow+\infty,  \nn\\
b&\rightarrow&b_0\equiv \arcsin{(\frac{1}{\sqrt{(1+\lambda\gamma^2)}})} \  \text{if}\  (\phi-\phi_o)\rightarrow0^+.
\ea
To study the asymptotics of the effective Friedmann and Raychaudhuri equations, we expand Eq.\eqref{Fri} and Eq.\eqref{Ray} by $b$ and $b-b_0$ respectively up to the second order term in the above large $v$ limit and thus obtain
\ba
H^2&=&\frac{8\pi G}{3}\rho, \quad\quad ( \mbox{for} \quad b\rightarrow 0)\label{asyFri}\\
H^2&=&\left(\frac{1-5\lambda\gamma^2}{1+\lambda\gamma^2}\right)\frac{8\pi G\rho}{3}+\frac{\Lambda}{3}, \quad(\mbox{for} \quad b\rightarrow b_0)\label{Friedmaneq}
\ea
and
\begin{equation}
\frac{\ddot{a}}{a}=-\frac{4\pi G}{3}(\rho+3P),\quad ( \mbox{for} \quad b\rightarrow 0)\label{asyRay},
\end{equation}
\begin{equation}
\frac{\ddot{a}}{a}=-\left(\frac{1-5\lambda\gamma^2}{1+\lambda\gamma^2}\right)\frac{4\pi G}{3}(\rho+3P)+\frac{\Lambda}{3},\quad (\mbox{for} \quad b\rightarrow b_0)\label{Rayeq},
\end{equation}
where we defined the pressure of matter by $P=-\frac{\partial H_\phi}{\partial V}$, and an effective cosmological constant \ba
\Lambda\equiv\frac{3\lambda}{(1+\lambda\gamma^2)^2\Delta}\label{effectivec}.
\ea The asymptotic behaviors of the scalar curvature $R=6(H^2+\frac{\ddot{a}}{a})$ are given by
 \begin{equation}
R=-16\pi G\rho,\quad (\mbox{for}\quad b\rightarrow 0),
\end{equation}
\begin{equation}
R=-16\pi G\rho\left(\frac{1-5\lambda\gamma^2}{1+\lambda\gamma^2}\right)+4\Lambda,\quad (\mbox{for}\quad b\rightarrow b_0).
\end{equation}
\begin{figure}[!htb]
		\includegraphics [width=0.50\textwidth]{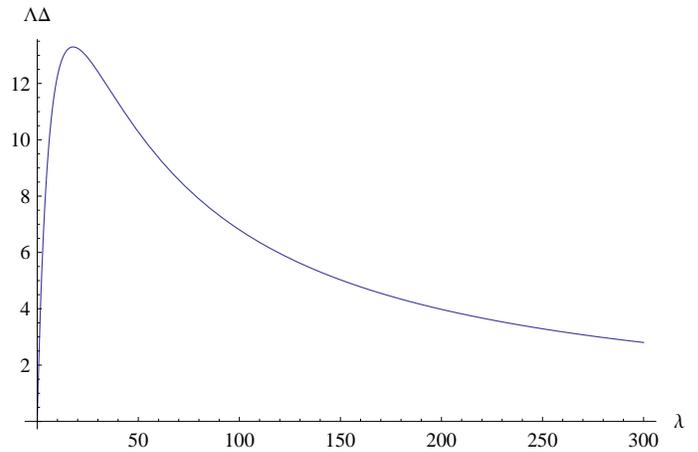}
		\caption{The $\Lambda\Delta$ as a function of $\lambda$. The $\gamma=0.2375$ is used in the calculation.}
		\label{fig:cosmologicalconstant}
	\end{figure}Thus, if the Universe started to collapse from an initial state of spatially flat FLRW configuration, it would undergo a quantum bounce and evolve into an asymptotic de-Sitter branch. A notable feature of this picture is that an effective cosmological constant
$\Lambda$ emerges from the quantum gravity effect. Eq.\eqref{Friedmaneq} implies an effective Newtonian constant \ba
G_\lambda=\frac{1-5\lambda\gamma^2}{1+\lambda\gamma^2} G\label{Glambda}
 \ea in the de-Sitter epoch. If G took the usual value of the Newtonian constant, for the special choice of $\lambda=1$, $G_\lambda$ would receive a large correction which is inconsistent with the current experiments \cite{Wang18}. However, the regularization freedom does admit us to choose certain sufficiently small $\lambda$ such that $G_\lambda$ satisfies the experimental restrictions.

The value of $\Lambda$ is determined by the value of $\lambda$ by Eq.\eqref{effectivec}, and the function $\Lambda\Delta(\lambda)$ is plotted by Fig.1. To reproduce the current observed cosmological constant $\Lambda_{ob}$, the corresponding $\lambda$ has two solutions, $\lambda_1\sim \frac{3}{\gamma^4\Delta\Lambda_{ob}}$ and $\lambda_2\sim\frac{\Delta\Lambda_{ob}}{3}$. It is obvious that for a reasonable choice of $\gamma\sim0.2$ coming from the calculation of black hole entropy\cite{Lewandowski04,Ash-view}, $\lambda_1$ is too big to give an acceptable $G_\lambda$. Moreover, in this case the critical density $\rho_{c}\sim\frac{3\Lambda_{ob}}{32\pi G}$ becomes very small and thus conflicts with the experiments.
However, $\lambda_2$ is sufficiently small to give an acceptable $G_\lambda$ and in the meantime leads to a bouncing density of Planck order as $\rho_{c}=\frac{3}{32\pi G(1+\lambda\gamma^2)\gamma^2\Delta}\sim\frac{3}{32\gamma^2\pi G\Delta}$. For such a $\lambda$, Eqs. \eqref{Friedmaneq} and \eqref{Rayeq} reduce to
\ba
H^2&=&\frac{8\pi G\rho}{3}+\frac{\Lambda_{ob}}{3},\\
\frac{\ddot{a}}{a}&=&-\frac{4\pi G}{3}(\rho+3P)+\frac{\Lambda_{ob}}{3}.
\ea They are nothing but the standard Friedmann and Raychaudhuri equations with the observational cosmological constant! Therefore, by choosing $\lambda=\frac{\Delta\Lambda_{ob}}{3}\thicksim 10^{-122}$, the standard dynamical equations can be obtained at large volume limit of the asympotic de-Sitter branch.

\section{Discussion}
Since the observational cosmological constant is so small, one may be worried about that the sub-leading quantum correction terms such as those in Eq. \eqref{expectationH} could influence the observational choice of the parameter $\lambda$. To check this issue, let us consider the effective Hamiltonian constraint of the model with sub-leading order \cite{YDM}
{\small\ba
H_F&=&-\frac{3\beta}{8\pi G\gamma^2\Delta}|v|\sin^2b\left(1-(1+\lambda\gamma^2)\sin^2(b)+2\epsilon^2\right)\nn\\
&&+\beta|v|\rho\left(1+\frac{1}{2\abs{v}^2\epsilon^2}+\frac{\hbar^2}{2\sigma^2p^2_{\phi}}\right)\label{effectiveH1},
\ea}where $\sigma$ denotes the "width" of the Gaussian semiclassical state of matter field. By assuming that the time derivatives of the quantum corrections are neglectable higher order terms. Straightforward calculations similar to those of Eqs. \eqref{Fri}, \eqref{Ray}, \eqref{asyFri}, \eqref{Friedmaneq}, \eqref{asyRay} and \eqref{Rayeq} give \ba
H^2&=&\frac{8\pi G}{3}\rho_{eff}-\frac{2\epsilon^2}{\gamma^2\Delta},\quad ( \mbox{for} \quad b\rightarrow 0)\\
H^2&=&\frac{8\pi G_\lambda}{3}\rho_{eff}+\frac{\Lambda}{3}\nn\\
&&-\left(\frac{1-5\lambda\gamma^2}{1+\lambda\gamma^2}\right)\frac{2\epsilon^2}{\gamma^2\Delta},\quad (\mbox{for} \quad b\rightarrow b_0)\\
\frac{\ddot{a}}{a}&=&-\frac{4\pi G}{3}(\rho_{eff}+3P)-\frac{2\epsilon^2}{\gamma^2\Delta},( \mbox{for} \quad b\rightarrow 0)\\
\frac{\ddot{a}}{a}&=&-\frac{4\pi G_\lambda}{3}(\rho_{eff}+3P)+\frac{\Lambda}{3}\nn\\
&&-\left(\frac{1-5\lambda\gamma^2}{1+\lambda\gamma^2}\right)\frac{2\epsilon^2}{\gamma^2\Delta},\quad (\mbox{for} \quad b\rightarrow b_0)\label{Rayeq1}
\ea where $\rho_{eff}=\rho\left(1+\frac{1}{2\abs{v}^2\epsilon^2}+\frac{\hbar^2}{2\sigma^2p^2_{\phi}}\right)$. Although the value of the quantum fluctuation or the Gaussian spread $\epsilon$ could not be determined in the model, one may always ask $\widetilde{\Lambda}=\Lambda-\left(\frac{1-5\lambda\gamma^2}{1+\lambda\gamma^2}\right)\frac{6\epsilon^2}{\gamma^2\Delta}$ to coincide with $\Lambda_{ob}$. Then , taking account of the observational restriction of $G_\lambda$ in \eqref{Glambda}, the  regularization parameter should be fixed as
\ba
\lambda&=&\frac{B-\sqrt{B^2-4(\Lambda_{ob}\Delta\gamma^2-30\epsilon^2)(\Lambda_{ob}\Delta\gamma^2+6\epsilon^2)}}{2(\Lambda_{ob}\Delta\gamma^2-30\epsilon^2)\gamma^2}\nn\\
&\approx&\frac{\Lambda_{ob}\Delta}{3}+\frac{2\epsilon^2}{\gamma^2}
\ea with $B=3+24\epsilon^2-2\Lambda_{ob}\Delta\gamma^2$. Therefore, even if the sub-leading quantum corrections were comparable with the observational cosmological constant, their effect could always be absorbed into the effect of the parameter $\lambda$.

To summarize, an one-parameter regularization freedom of the Hamiltonian constraint for LQG is introduced by Eq.\eqref{generalH}. The corresponding spatially flat FLRW model is studied. The quantum difference equation representing the evolution of the universe and its effective Hamiltonian are obtained. The general expression \eqref{generalH} includes the Euclidean term quantum dynamics \cite{APS3} by choosing $\lambda=-\frac{1}{\gamma^2}$ and the Euclidean-Lorentzian term dynamics \cite{YDM} by choosing $\lambda=1$. The quantum bounce nature of LQC is tenable in the general case when the matter density approaches certain critical density. For a chosen $\lambda>0$, the effective Hamiltonian of our LQC model can lead to a branch of Universe with an asympotic positive cosmological constant connecting to the FLRW branch through the quantum bounce. Remarkably, by suitable choice of $\lambda$, the standard Friedmann equation with the observational cosmological constant $\Lambda_{ob}$ can be obtained at large volume limit of the asymptotic de-Sitter branch. In the meantime, unlike the case of $\lambda=1$, the effective Newtonian constant $G_\lambda$ also satisfies the experimental restrictions. The significance of the current work is that we found a choice of regularization such that the acceleration expansion of our universe (dark energy) could be attributed to the emergent effects of quantum gravity. From the expression of \eqref{effectiveH}, there does exist an one-parameter ambiguity. This is related to the well-known fact that the classically equivalent expressions would generally be nonequivalent after quantization. By using the observational data, we successfully  single out the preferred value of $\lambda$ in the expression of the Hamiltonian. In addition, this effect is generated from dynamics of quantum gravity rather than a non-dynamical constant added by hand. Since the effective Hamiltonian \eqref{effectiveH} can also be derived from full LQG by suitable semiclassical analysis \cite{DL18,HL}, our result indicates a fair possibility to emerge the dark energy from LQG.

\section*{Acknowledgments}
This work is supported by NSFC with Grants No. 11775082, No. 12047519, No. 11961131013 and No. 11875006.


\end{document}